\documentclass[preprint,aps,amssymb,superscriptauthor,nofootinbib]{revtex4}

\usepackage{graphicx}

\begin{document}

\title{SU(2) gauge theory of gravity with topological invariants}

\author{Sandipan Sengupta}
\email{sandi@imsc.res.in}
\address{The Institute of Mathematical Sciences, Chennai-600113, India}

\newcommand{\del}{\partial}
\newcommand{\f}{\frac}
\newcommand{\Case}[2]{{\textstyle \frac{#1}{#2}}}
\newcommand{\lP}{\ell_{\mathrm P}}
\def\ee{\end{equation}}
\def\beq{\begin{eqnarray}}
\def\eeq{\end{eqnarray}}
\def\bn{\begin{eqnarray*}}
\def\en{\end{eqnarray*}}
\def\w{\omega}
\def\g{\gamma}
\def\S{\Sigma}
\def\s{\sigma}
\def\t{\tau}
\def\a{\alpha}
\def\b{\beta}
\def\m{\mu}
\def\n{\nu}
\def\D{\Delta}
\def\d{\delta}
\def\r{\rho}
\def\l{\lambda}
\def\th{\theta}
\def\k{\kappa}
\def\e{\epsilon}
\def\P{\Psi}
\def\he{\hat =}

\begin{abstract}
  The most general gravity Lagrangian in four dimensions contains
  three topological densities, namely Nieh-Yan, Pontryagin and Euler,
  in addition to the Hilbert-Palatini term. We set up a Hamiltonian
  formulation based on this Lagrangian. The resulting canonical theory
  depends on three parameters which are coefficients of these terms
  and is shown to admit a real $SU(2)$ gauge theoretic interpretation
  with a set of seven first-class constraints. Thus, in addition to
  the Newton's constant, the theory of gravity contains three
  (topological) coupling constants, which might have non-trivial
  imports in the quantum theory.
 
\end{abstract}
\maketitle
\section{Introduction}
The classical dynamics of a system is not affected by the addition of
topological densities in the Lagrangian. This is so because such
densities can always be locally written as total divergences. However,
quantum dynamics might depend on them. The cases of the Sine-Gordon
quantum mechanical model or QCD provide perfect examples of such a
phenomenon where topological terms leave their imprints on the quantum
theory\cite{coleman}.

In gravity theory in 3+1 dimensions, there are three possible
topological terms, namely, Nieh-Yan, Pontryagin and Euler, which can
be added to the Lagrangian. In terms of tetrads and spin-connections,
these can be written as\footnote{The quantity $\tilde{X}^{IJ}$ the dual
  of $X_{IJ}$ in the internal space:
  $\tilde{X}^{IJ}=\frac{1}{2}~\epsilon^{IJKL}X_{KL}$}: \beq I_{NY}^{} &~=&
~e\Sigma^{\m\n}_{IJ} {\tilde R}^{~~~IJ}_{\m\n}(\w) ~+ ~\e^{\m\n\a\b}
D_\m(\w)e_{I\n} D_\a(\w)e^I_\b~=~\partial_\m
\left[ \e^{\m\n\a\b} ~e^I_\n~ D_\a(\w) e_{I\b} \right]\nonumber\\
I_P^{} &~= &~\e^{\m\n\a\b} R_{\m\n IJ} (\w) R_{\a\b}^{~~~IJ}(\w)
~=~4\partial_\m \left[\e^{\m\n\a\b} \w^{~IJ}_\n \left(
    \partial_\a \w_{\b IJ} + {\frac 2 3} \w^{~~~K}_{\a I} \w_{\b KJ}
  \right)\right] \nonumber\\
I_E^{} &~=&~ \e^{\m\n\a\b} R_{\m\n IJ} (\w) {\tilde
  R}_{\a\b}^{~~~IJ}(\w) ~=~ 4\partial_\m \left[\e^{\m\n\a\b}
  {\tilde\w}^{~IJ}_\n \left(\partial_\a \w_{\b IJ} + {\frac 2 3}
    \w^{~~~K}_{\a I} \w_{\b KJ} \right)\right]~~~~\eeq where
$R^{~IJ}_{\mu\nu}(\omega) = \del_{[\mu} \omega_{\nu]}^{~IJ}
+\omega_{[\mu}^{~IK} \omega_{\nu]K}^{~~~J}$ and
$D_{\mu}(\omega)e_{\nu}^{I}=\del_{\mu}e_{\nu}^{I}+\omega_{\mu}^{~IJ}e_{\nu
  J}$. Although these topological densities are functions of local
geometric quantities, they encode only the global properties of the
manifold. The Nieh-Yan density depends on torsion and in Euclidean
theory, its integral over a compact manifold is a sum of three
integers associated with the homotopy maps $\pi_3(SO(5))=Z$ and
$\pi_3(SO(4))=Z+Z$. Among the other two which depend on the curvature,
the Pontryagin-class characterises the integers corresponding to the
map $\pi_3(SO(4))=Z+Z$ and the Euler-class characterises the
combination of Betti numbers.  While the first two densities are P and
T odd, the third is P and T even (see \cite{kaul1} and the references
within).

In order to understand their possible import in the quantum theory, it
is important to set up a classical Hamiltonian formulation of the
theory containing all these terms in the action.  In ref.\cite{date},
such an analysis has been presented for a theory based on Lagrangian
density containing the standard Hilbert-Palatini term and the Nieh-Yan
density. The resulting theory, in time gauge, has been shown to
correspond to the well-known canonical gauge theoretic formulation of
gravity based on Sen-Ashtekar-Barbero-Immirzi {\it real} $SU(2)$ gauge
fields \cite{ashtekar}. Here $\eta^{-1}$, the inverse of the
coefficient of Nieh-Yan term, is identified with the Barbero-Immirzi
parameter $\gamma$.  The framework in \cite{date} supersedes the
earlier formulation of Holst \cite{holst} in the sense that unlike the Holst term, the Nieh-Yan density \\
(a) does not need any further modifications for the inclusion of
matter couplings and the equations of motion continue to be
independent
of $\eta$ for all couplings;\\
(b) provides a topological interpretation for Barbero-Immirzi
parameter, leading to a
complete analogy between $\eta$ and the $\theta$-parameter of non-abelian gauge theories (from the classical perspective). \\
As an elucidation of these facts, the method has been applied to
spin-$\frac{1}{2}$ fermions\cite{date} and supergravity
theories\cite{sengupta}.

Here we include all three topological terms in the Hilbert-Palatini
Lagrangian\cite{kaul1}: \beq {\cal L}(e,\omega) ~=~ {\frac 1 2}~ e~ \Sigma^{\m\n}_{IJ}
~ R^{~~~IJ}_{\m\n}(\w) ~+~ {\frac \eta 2} ~I^{}_{NY} ~+~ {\frac \th 4}
~I^{}_P ~+ ~{\frac \phi 4} ~I^{}_E
\label{L1} \eeq where, $ \Sigma^{\mu\nu}_{IJ}=
\frac{1}{2}~e^{\mu}_{[I} e^{\nu}_{J]}$. In order to understand how the
canonical theory of gravity gets affected by such additions, a
Hamiltonian analysis based on this Lagrangian is presented below,
demonstrating how we obtain a real SU(2) formulation of gravity with
all three topological densities.

\section{Hamiltonian formulation}
A convenient way to proceed is to decompose the 16 tetrad fields
$e^{I}_{\mu}$ into the fields $V_{a}^{I}$, $M_{I}$, $N^{a}$ and $N$
(16=9+3+3+1) (see \cite{kaul1} for further details): \beq e^{I}_{t} &
=& N M^{I}+N^{a}V_{a}^{I}, ~~~~~ e^{I}_{a} =
V^{I}_{a}~; \nonumber \\
e^{t}_{I} & =& -\frac{M_{I}}{N}~, ~~~~~ e^{a}_{I} =
V^{a}_{I}+\frac{N^{a}M_{I}}{N};\nonumber\\
M_{I}V_{a}^{I} &= &0~~,~~~~~ M_{I}M^{I} = -1~; \nonumber \\
V_a^I V^b_I &= &\delta_a^b~, ~~~~ V_a^I V^a_J = \delta^I_J + M^IM_J~.
\label{tetpara2} \eeq Next, instead of the variables $V^{a}_{I}$ and
$M^{I}$, we define a new set of 12 variables as: \beq ~E^{a}_{i} &=&
2e\Sigma^{ta}_{0i}~\equiv~ e\left(e^t_0 e^a_i - e^t_i e^a_0\right)= -
~ {\sqrt q}~ M^{}_{[0} V^a_{i]}, ~~~~~~~~\chi_{i} = -M_{i}/M^{0}~ ~~~~
\label{E1} \eeq 

Before writing the full Lagrangian, we note that with the help of
Bianchi identities $\e^{abc}_{} D^{}_a(\w) R^{}_{bcIJ} = 0$ and
$\e^{abc}_{} D^{}_a(\w) \tilde{R}^{}_{bcIJ} = 0$, the last two terms
in (\ref{L1}) can be written as\footnote{The quantity $X^{(\eta)IJ}$
  is defined as: $X^{(\eta)IJ}=X^{IJ}+\eta
  ~\tilde{X}^{IJ}$}:\beq\frac{\th}{4} ~I_P^{}+ \frac{\phi}{4} ~I_E^{}
= e^a_{IJ} ~ \partial_t^{} \w^{(\eta)IJ}_a\label{PE}\eeq with
$\left(1+\eta^2_{}\right) e^a_{IJ} ~ = ~ \e^{abc}_{}
\left\{ \left(\th+\eta \phi\right) R^{}_{bcIJ}(\w) + \left( \phi -
    \eta \th \right) {\tilde R}^{}_{bcIJ}(\w) \right\}\label{eIJ}$.
Using (\ref{tetpara2}), (\ref{E1}) and (\ref{PE}), the Lagrangian in
(\ref{L1}) can be written as: \beq {\cal
  L}~=~\pi^{a}_{IJ}\del_{t}\omega_{a}^{(\eta)IJ}~+~t^{a}_{I}\del_{t}V_{a}^{I}~-~NH~-~N^{a}H_{a}~-~\frac{1}{2}\omega_{t}^{IJ}G_{IJ}
\label{L4} \eeq where $\pi^{a}_{IJ}~=~e~\Sigma^{ta}_{IJ}~+~e^{a}_{IJ}$
and \beq
G_{IJ}~&=&~   -2D_{a}(\w)\pi^{a(\eta) }_{IJ} -t^a_{[I}V^{}_{J]a}~~~, \nonumber\\
H_{a}~&=&~
\pi^{b}_{IJ}R_{ab}^{(\eta)IJ}(\w)~-~V_{a}^{I}D_{b}(\omega)t^{b}_{I}~~~, \nonumber\\
H~&=&~\frac{2 }{{\sqrt q}~} \left( \pi^{a(\eta) }_{IK} -e^{a(\eta) }
  _{IK}\right)\left(\pi^{b(\eta) }_{JL} -e^{b(\eta)}_{JL}\right)
\eta^{KL}R_{ab}^{~~ IJ}(\omega) -M^{I}D_{a}(\omega)t^{a}_{I} ~~~~.
\label{H2} \eeq Since there are no velocities associated with the
fields $N, ~N^a$ and $\omega_{t}^{IJ}$, we have the constraints
$H\approx 0,~H_a \approx 0,~G_{IJ}\approx 0~~$.

Next, we split the 18 spin-connection fields $\omega_{a}^{IJ}$ as:
\beq A^i_a~\equiv~ \w^{(\eta)0i}_a =\w^{0i}_a +\eta{\tilde
  {\w}}^{0i}_a, ~~~~~~~K^i_a ~\equiv~ \w^{0i}_a~~.\eeq The rationale
behind such a choice is to make the SU(2) interpretation transparent,
as can be understood by noting that $A^i_a$ transforms as connection
and $K_a^i$ as adjoint representation under the SU(2) gauge
transformations.  Also, it is convenient (although not necessary) to
work in the time gauge where the boost constraints are solved by the
gauge choice $\chi_i=0$.  Thus, in this gauge, the symplectic form
becomes: \beq
\pi^{a}_{IJ}\del_{t}\omega_{a}^{(\eta)IJ}~+~t^{a}_{I}\del_{t}V_{a}^{I}~
~=~\hat{E}^{a}_{i}\del_{t}A_{a}^{i}~+~\hat{F}^{a}_{i}\del_{t}K_{a}^{i}~+~t^{a}_{i}\del_{t}V_{a}^{i}\label{symp}
\eeq with
 \begin{eqnarray} 
 {\hat E}^i_a ~&\equiv&~-~\frac{2}{\eta}
  ~\tilde{\pi}^{a(\eta) }_{0i}~\equiv~ -~\frac{2}{\eta}
 ~ \left( {\tilde \pi}^a_{0i}- \eta \pi^a_{0i} \right) ~=~ E^a_i   -\frac{2}{\eta}~{\tilde e}^{a(\eta) }_{0i}(A,K)  
  \\
  {\hat F}^a_i ~&\equiv&~2\left(\eta
    +\frac{1}{\eta}\right)\tilde{\pi}^{a}_{0i}~=~ 2\left( \eta +
    \frac{1}{\eta} \right) 
  {\tilde e}^a_{0i}(A,K)\label{F}
\end{eqnarray}
Here, the fields $V^i_a$ and its conjugate $t^{a}_{i}$ are not
independent; they obey the following second-class constraints: \beq
V^i_a ~- ~\frac{1}{\sqrt E}~E^i_a ~\equiv~ 0~, ~~~~~~~~t^a_i ~-~ \eta
\e^{abc}_{} D_b(\w) V^i_c ~= ~ \e^{abc}_{} \left( \eta D_b(A) V^i_c -
  \e^{ijk}_{} K^j_b V^k_c \right) ~ \label{tau3} \eeq Similarly,
eq.(\ref{F}) shows that the momenta ${\hat F}^{a}_{i}$ obey
constraints of the form\beq \chi^{a}_{i}~:= ~{\hat
  F}^{a}_{i}-f(A_b^j,K_c^k)~\approx~0\label{F1}\eeq These imply
secondary constraints: \beq \left[ \chi^a_i(x), H(y) \right]
~\approx~0 ~=>~ t^a_i - \left(\frac{1+\eta^2_{}}{\eta^2_{}}
\right)\left\{\eta\e^{ijk}_{} D^{}_b(A) \left({\sqrt E} E^a_jE^b_k
  \right) +{\sqrt E}E^{[a}_j E^{b]}_i K^j_b \right\}~ \approx~ 0
~\label{t}\eeq The solution of (\ref{t}) can be expressed in the form:
$K^i_a-\kappa^i_a(A_b^j,E^c_k)\approx 0~$.  Since $K^i_a$ and $\hat{F}^{a}_{i}$ are canonically conjugate, these constraints evidently
form a second-class pair with (\ref{F1}). The constraints
(\ref{tau3}), (\ref{F1}) and (\ref{t}), alongwith the constraints
$G_{i}^{rot} \approx 0,~ H_a \approx 0 ~, ~H \approx 0 $, completely
characterise the canonical theory corresponding to the Lagrangian
density (\ref{L1}).

Notice that for $\theta=0$ and $\phi=0$, the momenta $\hat{F}^{a}_{i}$
in (\ref{F}) vanishes. This corresponds to the Barbero-Immirzi
formulation. Thus, the effect of the addition of Pontryagin and Euler
terms in the Lagrangian gets reflected through a richer symplectic
structure characterised by a non-vanishing $\hat{F}^{a}_{i}$. Also,
for non-vanishing $\theta$ and $\phi$, the canonical conjugate of the
connection $A_{a}^{i}$ is $\hat{E}^{a}_{i}$, and not the densitized
triad $E^{a}_{i}$ as in the case for $\theta=0,~\phi=0$.

\section{SU(2) interpretation}
The second-class constraints can all be implemented by using the
corresponding Dirac brackets instead of the Poisson brackets. 
After imposing all the second-class pairs strongly, we are left with a
set of seven first class constraints: \beq
&& G^{rot}_i  ~\equiv~  \eta ~D^{}_a(A){\hat E}^a_i ~+~\e^{ijk}_{} K^j_a {\hat F}^a_k ~\approx~ 0 \nonumber \\
&& H^{}_a ~\equiv~ {\hat E}^b_i F^i_{ab}(A) ~+~ {\hat F}^b_i
D^{}_{[a}(A) K^i_{b]} ~- ~K^i_a D^{}_b(A) {\hat F}^b_i
- {\eta}^{-1}_{} ~G^{rot}_i K^i_a ~\approx~ 0 \nonumber \\
&& H ~\equiv~ \frac{\sqrt E}{2\eta} \e^{ijk}_{} E^a_i E^b_j
F^k_{ab}(A) -\left( \frac{1+\eta^2_{}}{2\eta^2_{}}\right){\sqrt E}
E^a_i E^b_j K^i_{[a} K^j_{b]} +
\frac{1}{\eta}~\partial^{}_a\left({\sqrt E} G^{rot}_k E^a_k\right)
\approx ~0 ~~~~~~~~~\label{finalconstraints} \eeq Evaluating the Dirac
brackets of the rotation constraints $G^{rot}_{i}$ with the basic
fields, we find that they are the generators of the SU(2)
gauge transformations: \beq
\left[G^{rot}_i(x), ~{\hat E}^a_j(y)\right]_D^{} &=& \e^{ijk}_{}{\hat E}^a_k ~\delta^{(3)}_{}(x,y)~, \nonumber \\
\left[G^{rot}_i (x), ~A^j_a (y)\right]_D^{} & = & - \eta \left(
  \delta^{ij}_{} \partial_a^{} ~ + {\eta}^{-1}_{}~\e^{ikj}_{} A^k_a
\right) ~\delta^{(3)}_{}(x,y)~~. \eeq

Thus, we have a SU(2) gauge theory of gravity with all three
topological parameters. The Barbero-Immirzi parameter $\eta^{-1}$ acts
as the coupling constant of gauge field $A_{a}^{i}$ whereas the other
two parameters $\theta,\phi$ enter in the definition of its conjugate
$\hat{E}^{a}_{i}$. Since the topological densities are all functions
of the geometric fields (i.e. tetrads and spin-connections), addition of
matter coupling does not affect such a gauge theoretic interpretation of
gravity.

\section{Concluding remarks}
It is important to investigate the imports of these topological terms
in the quantum theory of gravity. Although the Barbero-Immirzi
parameter is known to appear in the area spectrum in Loop Quantum
Gravity, the role of the other two parameters in quantum geometry is
yet to be understood. Also, the question that whether these terms
imply non-trivial topological sectors and potential instanton effects
in the quantum theory similar to the non-abelian gauge theories
demands a detailed study (for a relevant discussion, see
\cite{sengupta1} and the references within).

As a final remark, we note that the Dirac bracket between $A_{a}^{i}$
and $\hat{E}^{a}_{i}$ is not a canonical one, unlike the
Barbero-Immirzi formulation. Although this need not be an issue as far
as the classical theory is concerned, quantization based on these
canonical variables is not straightforward. However, as demonstrated
in \cite{kaul1}, it is possible to find a suitable canonical pair
which leads to the standard bracket, thus providing a smooth passage
towards the quantum theory.
\vspace{.1cm}\\

{\bf Acknowledgments\\} 
The author thanks Romesh Kaul for collaboration on this topic and the
organisers of Loops-11 for the wonderful hospitality at CSIC, Madrid,
where this work was presented.


\begin{thebibliography}{99}

\bibitem{coleman} S. Coleman, {\it Aspects of Symmetry}   (Cambridge University Press, 1985);\\
  R. Rajaraman, {\it Solitons and Instantons} (North-Holland, The
  Netherlands, 1982)
%
\bibitem{kaul1}	R. K. Kaul, S. Sengupta, arXiv:{\bf 1106.3027} [gr-qc] (2011)
%
\bibitem{date} G. Date, R.K. Kaul, S. Sengupta, Phys. Rev. {\bf D79}, 044008 (2009)
%
\bibitem{ashtekar} A. Ashtekar and J. Lewandowski, Class. Quant. Grav.
  {\bf 21}, R53 (2004)
%
\bibitem{holst} S. Holst, Phys. Rev. {\bf D53}, 5966-5969 (1996)
%
\bibitem{sengupta} S. Sengupta and  R.K. Kaul, Phys. Rev. {\bf D81}, 024024 (2010)
%
\bibitem{sengupta1} S. Sengupta, Class. Quantum Grav. {\bf 27}, 145008 (2010)
%
\end{thebibliography}
\end{document}